\documentclass[a4paper]{jpconf}
\usepackage{graphicx}
\usepackage{epsfig}		
\usepackage{dcolumn}
\usepackage{bm}
\usepackage{subfigure}

\usepackage[all]{xy}
\usepackage{amsmath,upgreek}
\usepackage{amssymb}

\usepackage{color}
\usepackage[colorlinks=true,
 linkcolor=red,
 urlcolor=purple,
 citecolor=blue,hyperindex]{hyperref}

\def\0{\mbox{\tiny $0$}}
\def\1{\mbox{\tiny $1$}}
\def\2{\mbox{\tiny $2$}}
\def\3{\mbox{\tiny $3$}}
\def\4{\mbox{\tiny $4$}}
\def\5{\mbox{\tiny $5$}}
\def\6{\mbox{\tiny $6$}}
\def\7{\mbox{\tiny $7$}}
\def\8{\mbox{\tiny $8$}}
\def\9{\mbox{\tiny $9$}}

\def\f14{\mbox{\tiny $\frac{1}{4}$}}

\def\bb#1{\mbox{\footnotesize $(#1)$}}

\begin{document}
\title{Anharmonic effects on phase-space quantum profiles: an exact approach}
\author{Alex E Bernardini and Caio Fernando e Silva}
\ead{alexeb@ufscar.br}
\address{Departamento de F\'{\i}sica, Universidade Federal de S\~ao Carlos, PO Box 676, 13565-905, S\~ao Carlos, SP, Brasil.}

\begin{abstract}
Given its well known spectral decomposition profile, the $1$-dim harmonic oscillator potential modified by an inverse square ($1$-dim angular momentum-like) contribution works as an efficient platform for probing classical and quantum information quantifiers in the context of the phase-space Weyl-Wigner formalism.
In particular, the phase-space informational content related to the canonical ensemble driven by such a {\em singular oscillator} can be quantified in terms of well established analytical structures.
Considering that, on one hand, the {\em singular oscillator} produce a spectral decomposition profile equivalent to that one of the unmodified harmonic system -- in the sense that they result into identical thermodynamic statistics, even for different statistical mixtures -- on the other hand, a more complete scrutinization of their phase-space information content can capture some different aspects of the encoded information for the related quantum ensembles.
Besides the identification of decoherence effects, the Wigner flow analysis is presumedly useful in identifying stable quantum configurations, according to finite temperature and interaction parameter values.
Unexpectedly, our results show that the equivalence between the statistical (quantum) mechanics of the anharmonic singular oscillator and an ordinary harmonic oscillator can also be extended to the phase-space quantum purity quantifier, which is analytically computed and reproduces exactly the same quantum ensemble statistical mixture profile, which does not depend on interaction parameter values.
\end{abstract}

\date{\today}

\section{Introduction}

\hspace{1 em} Given the role of the partition function in describing statistical properties, physical systems with similar spectral decomposition profiles can sometimes exhibit identical properties at thermodynamic equilibrium.
That is the case \cite{0001,0002} of the harmonic oscillator potential modified by the addition of an inverse square $1$-dim angular momentum-like contribution -- also approximated by a $1$-dim Coulomb-like dipole momentum contribution -- which thermodynamically produces the same quantum statistics of the $1$-dim harmonic ensemble.
Such a {\em singular oscillator} has its dynamics quantum mechanically described by an exactly solvable Schr\"odinger equation \cite{0003,0004,0005}, which provides, for instance, the setup for perturbative expansions \cite{0009,0010} and for the construction of solvable models of $N$-interacting systems \cite{0007,0008}.

Considering that the {\em singular oscillator} quantum description can be extended to the phase-space, their corresponding Wigner functions have been relevant in describing the interplay between quantum and classical aspects as it has been reported in some preliminary analysis \cite{NovoPaper,JCAP18}.
In fact, the Wigner formalism \cite{Wigner} encompasses the phase-space dynamics so as to include quantum effects and to provide a more complete overview of the quantum information encoded by quantum systems.
Since the foundations of quantum mechanics remain unaffected, in particular, when compared to a typical Schr\"odinger analysis, the phase-space Wigner formulation of quantum mechanics \cite{01A,02A,03A,Case} -- due to its natural reducibility to the classical Liouvillian formulation -- is supposed clear up the understanding of classical and quantum boundaries.

Moreover, the Wigner formalism can be connected with quantum statistical mechanics \cite{Hillery}.
In fact, the concept of statistical equilibrium described by the phase-space Gibbs-Boltzmann formula is not trivially generalized to quantum mechanics.
Even if generalizations are accepted, in quantum theory, there does not exist a simplified irrefutable expression for
the statistical probability that describe quantum ensembles: one cannot ask for the simultaneous probabilities for
the phase-space coordinates.
The thermodynamics of quantum mechanical systems brings up a set of encoded information when it is described by Wigner functions which, of course, can be used to straightforwardly calculate the corresponding quantum version of partition functions.
In this context, the proposal of this manuscript is to describe the complete phase-space profile of the above-mentioned anharmonic quantum system so as to construct canonical ensembles from which decoherence effects can be identified.
Therefore, for the so-called {\em singular oscillator}, obtaining the Wigner function for the corresponding quantum ensemble is supposed to be helpful in identifying (de)coherent thermodynamic configurations, according to finite temperature and interaction parameter values.

The manuscript is organized as follows.
In Section II, the phase-space information flux associated to the Wigner purity quantifier is recovered from the fluid analog description of the Wigner formalism \cite{EPL18,NovoPaper}.
Prerogatives and advantages of a dimensionless framework are established so as to simplify outcoming analysis.
Preliminary results for the Wigner function and Wigner currents for the $1$-dim harmonic oscillator plus inverse square potential are considered in Section III, where an overview of the analytical strategy for computing the quantum purity is exemplified for a typical quantum pure state.
The construction of canonical ensembles as thermalized statistical mixtures through the Wigner formalism is presented in Section IV. Straightforward results for the partition function and for measures of quantum purity are finally obtained and the identification of decoherence effects in terms of finite temperatures is provided.
Our conclusions are drawn in Section V and suggest an interpretative view of how our results are related with the decoherence profile of the Universe in the scope of predictions from quantum and classical cosmology.

\section{Wigner flux and quantum purity}

\hspace{1 em} Through the Weyl transform of a quantum operator, $\hat{O}$, 
\begin{equation}
O^W(q, p)
= 2\hspace{-.2cm} \int^{+\infty}_{-\infty} \hspace{-.15cm}du\,\exp{\left(2\,i \,p\, u/\hbar\right)}\,\langle q - u | \hat{O} | q + u \rangle= 2 \hspace{-.2cm} \int^{+\infty}_{-\infty} \hspace{-.15cm} dv \,\exp{\left(-2\, i \,q\, v/\hbar\right)}\,\langle p - v | \hat{O} | p + v\rangle,
\end{equation}
where $q$ and $p$ are position and momentum coordinates, respectively, one has the inception of the Wigner function, $W(q, p)$, which is identified by $O^W(q, p)$ if $\hat{O}$ is replaced by a quantum mechanical density matrix operator, $\hat{\rho}$, such that
\begin{equation}
 h^{-1} \hat{\rho} \to W(q, p) = (\pi\hbar)^{-1} 
\int^{+\infty}_{-\infty} \hspace{-.15cm}du\,\exp{\left(2\, i \, p \,u/\hbar\right)}\,
\psi^{\ast}(q - u)\,\psi(q + u),
\end{equation}
for the case of $\hat{\rho}$ identified by a pure state, $|\psi \rangle \langle \psi |$, and which has the properties of a real-valued {\em quasi}-probability distribution.

The connection of the Wigner function with the matrix operator quantum mechanics is implemented according to the trace of the product between $\hat{\rho}$ and a generic operator, $\hat{O}$, evaluated through the integral of the product of their Weyl transforms over the infinite volume described by the phase-space coordinates, $q$ and $p$, as \cite{Wigner,Case}
\begin{equation}
Tr_{\{q,p\}}\left[\hat{\rho}\hat{O}\right] \to \langle O \rangle = 
\int^{+\infty}_{-\infty} \hspace{-.15cm}\int^{+\infty}_{-\infty} \hspace{-.15cm} {dq\,dp}\,W(q, p)\,{O^W}(q, p).
\label{five}
\end{equation}
It suggests a consistent probability distribution interpretation which is constrained by the normalization condition over $\hat{\rho}$, $Tr_{\{q,p\}}[\hat{\rho}]=1$.

Generically speaking, the matrix operator quantum mechanics aspects reproduced by the Weyl-Wigner formalism are even more evinced when the statistical aspects related to the nature of the density matrix are noticed: the Weyl-Wigner formalism also admits extensions from pure states to statistical mixtures, for which the quantum purity is computed through an analogous of the trace operation, $Tr_{\{q,p\}}[\hat{\rho}^2]$, read as
\begin{equation}
Tr_{\{q,p\}}[\hat{\rho}^2] = 2\pi\hbar\int^{+\infty}_{-\infty} \hspace{-.15cm}\int^{+\infty}_{-\infty} \hspace{-.15cm} {dq\,dp}\,W(q, p)^2,
\label{pureza}
\end{equation}
with the factor $2\pi\hbar$ introduced to satisfy the pure state constraint: $Tr_{\{q,p\}}[\hat{\rho}^2] = Tr_{\{q,p\}}[\hat{\rho}] = 1$. 

A completer overview of the phase-space Weyl-Wigner formalism is described by its related flow field, $\mathbf{J}(q,\,p;\,t)$, which describes a phase-space fluid-analog of the Wigner function \cite{02A,Ferraro11,Donoso12,04}.
The vector flux, $\mathbf{J}(q,\,p;\,t)$, is decomposed into the phase-space coordinates, $q$ and $p$, that is $\mathbf{J} = J_q\,\hat{q} + J_p\,\hat{p}$, in order to return the continuity equation given by \cite{Case,Ballentine,02A,EPL18}
\begin{equation}
\frac{\partial W}{\partial t} + \frac{\partial J_q}{\partial q}+\frac{\partial J_p}{\partial p} \equiv
\frac{\partial W}{\partial t} + \mbox{\boldmath $\nabla$}\cdot \mathbf{J} =0.
\label{alexquaz51}
\end{equation}
For a non-relativistic quantum Hamiltonian written in the matrix operator representation as
\begin{equation}
H(\hat{Q},\,\hat{P}) = \frac{\hat{P}^2}{2m} + V(\hat{Q}),
\end{equation}
where $ V(\hat{Q})$ is the potential which shall be identified with the so-called {\em singular oscillator}, and which, from  the Weyl transform, returns $H(\hat{Q},\,\hat{P}) \to H^{W}(q,p)$, one has \cite{NovoPaper}
\begin{equation}
J_q(q,\,p;\,t)= \frac{p}{m}\,W(q,\,p;\,t), \label{alexquaz500BB}
\end{equation}
and
\begin{equation}
J_p(q,\,p;\,t) = -\sum_{\eta=0}^{\infty} \left(\frac{i\,\hbar}{2}\right)^{2\eta}\frac{1}{(2\eta+1)!} \, \left[\left(\frac{\partial~}{\partial q}\right)^{2\eta+1}\hspace{-.5cm}V(q)\right]\,\left(\frac{\partial~}{\partial p}\right)^{2\eta}\hspace{-.3cm}W(q,\,p;\,t).
\label{alexquaz500}
\end{equation}
The contributions from $\eta \geq 1$ in the above series expansion describe the quantum corrections which distort the classical Liouvillian trajectories. The suppression of such contributions, once applied into the Eq.~\eqref{alexquaz51}, yields the classical profile reproduced by the Liouville equation.

By identifying a mass scale through the parameter $m$, and introducing an energy scale given by $\hbar \omega$, where $\hbar$ is the Planck constant, and $\omega$ is an arbitrary angular frequency, a simplifying overview of the problem can be achieved for a dimensionless version of $H^{W}(q,\,p)$, $\mathcal{H}(x,\,k) = k^2/2 + \mathcal{U}(x)$, i.e. where $\mathcal{H}(x,\,k)$ is written in terms of dimensionless variables, $x = \left(m\,\omega\,\hbar^{-1}\right)^{1/2} q$ and $k = \left(m\,\omega\,\hbar\right)^{-1/2}p$, such that $\mathcal{H} = (\hbar \omega)^{-1} H$ and $\mathcal{U}(x) = (\hbar \omega)^{-1} V\left(\left(m\,\omega\,\hbar^{-1}\right)^{-1/2}x\right)$.
In this case, the Wigner function is more conveniently cast in the dimensionless form of $\mathcal{W}(x, \, k;\,\omega t) \equiv  \hbar\, W(q,\,p;\,t)$, where $\hbar$ is absorbed by the phase-space volume integrations, with re-defined Wigner currents, $\mathcal{J}_x(x, \, k;\,\omega t)$ and $\mathcal{J}_k(x, \, k;\,\omega t)$,  satisfying, $\omega\, \partial_x\mathcal{J}_x \equiv  \hbar\, \partial_q J_q(q,\,p;\,t)$ and $\omega \,\partial_k\mathcal{J}_k\equiv  \hbar \,\partial_p J_p(q,\,p;\,t)$, so that they all can be recast in the form of \cite{NovoPaper}\footnote{Notice that the correspondence between $\phi(x,\,\tau)$ and $\psi(q;\,t)$ is consistent with their normalization constraints,
\begin{equation}
\int^{+\infty}_{-\infty} \hspace{-.2 cm}{dx}\,\vert\phi(x;\,\tau)\vert^2 =\int^{+\infty}_{-\infty}\hspace{-.2 cm}{dq}\,\vert\psi(q;\,t)\vert^2 = 1.
\end{equation}}
\small\begin{eqnarray}\label{alexDimW}
\mathcal{W}(x, \, k;\,\tau) &=& \pi^{-1} \int^{+\infty}_{-\infty} \hspace{-.15cm}dy\,\exp{\left(2\, i \, k \,y\right)}\,\phi(x - y;\,\tau)\,\phi^{\ast}(x + y;\,\tau),\quad \mbox{with $y = \left(m\,\omega\,\hbar^{-1}\right)^{1/2} u$},\,\,\,\,\\
\label{alexDimWA}\mathcal{J}_x(x, \, k;\,\tau) &=& k\,\mathcal{W}(x, \, k;\,\tau)
,\\
\label{alexDimWB}\mathcal{J}_k(x, \, k;\,\tau) &=& -\sum_{\eta=0}^{\infty} \left(\frac{i}{2}\right)^{2\eta}\frac{1}{(2\eta+1)!} \, \left[\left(\frac{\partial~}{\partial x}\right)^{2\eta+1}\hspace{-.5cm}\mathcal{U}(x)\right]\,\left(\frac{\partial~}{\partial k}\right)^{2\eta}\hspace{-.3cm}\mathcal{W}(x, \, k;\,\tau),
\end{eqnarray}\normalsize
where $\tau = \omega t$ is also a dimensionless quantity, and from which one re-obtains the (dimensionless) continuity equation in terms of the phase-space coordinate vector, $\mbox{\boldmath $\xi$} = (x,\,k)$, as
\begin{equation}
\frac{\partial \mathcal{W}}{\partial \tau} + \frac{\partial \mathcal{J}_x}{\partial x}+\frac{\partial \mathcal{J}_k}{\partial k} = \frac{\partial \mathcal{W}}{\partial \tau} + \mbox{\boldmath $\nabla$}_{\xi}\cdot\mbox{\boldmath $\mathcal{J}$} =0.
\end{equation}
Through a similar procedure, a self-contained framework for obtaining additional continuity equations for the phase-space information flux properties of the Wigner functions can be straightforwardly devised \cite{NovoPaper,JCAP18,EPL18}.
Even considering that some relevant results concerned with von Neumann and R\'enyi entropy definitions have been discussed in some preliminary issues \cite{EPL18,NovoPaper02}, our analysis will be restricted to only formal aspects related to the quantum purity of quantum systems.

Firstly, considering that phase-space points, $\mbox{\boldmath $\xi$}$, are surrounded by an infinitesimal volume quantity, $dV \equiv dx\,dk$, one notices that a {\em substantial derivative} \cite{02A,EPL18} can be implemented by 
\begin{equation}
\frac{D~}{D\tau} \int_{V}dV\,\mathcal{W} \equiv 
\int_{V}dV\,\left[\frac{D\mathcal{W}}{D\tau} + \mathcal{W} \mbox{\boldmath $\nabla$}_{\xi}\cdot \mathbf{v}_{\xi}\right]\label{alexquaz57D},
\end{equation}
where $ \mathbf{v}_{\xi}$ corresponds to the classical phase-space vector velocity.
For quantifying the flux of information, a two-dimensional comoving closed surface, $\mathcal{C}$, is depicted by the trajectory obtained from the classical velocity, $\mathbf{v}_{\xi(\mathcal{C})} = (k,\, -\partial \mathcal{U}/\partial x)$, through which one has
\begin{equation}
\frac{D \mathcal{W}}{D\tau} = - \mathcal{W}\, \mbox{\boldmath $\nabla$}_{\xi} \cdot \mathbf{v}_{\xi(\mathcal{C})},
\label{alexquaz57B}
\end{equation}
for the classical Liouvillian limit of the flow field. For ${D \mathcal{W}}/{D\tau} = 0$, one has a conservation law which is translated into a divergenceless behavior of the fluid-analog expressed by $\mbox{\boldmath $\nabla$}_{\xi} \cdot \mathbf{v}_{\xi(\mathcal{C})} = 0$.

To identify the quantum distortions, one parameterizes the Wigner vector current as $\mbox{\boldmath$\mathcal{J}$} = \mathbf{w}\,\mathcal{W}$, where such a non-classical (pseudo)velocity, $\mathbf{w}$, satisfies the following constraint equation,
\begin{equation}
\mbox{\boldmath $\nabla$}_{\xi} \cdot \mathbf{w} = \frac{\mathcal{W}\, \mbox{\boldmath $\nabla$}_{\xi}\cdot \mbox{\boldmath$\mathcal{J}$} - \mbox{\boldmath$\mathcal{J}$}\cdot\mbox{\boldmath $\nabla$}_{\xi}\mathcal{W}}{\mathcal{W}^2} \neq0,
\label{alexquaz59}
\end{equation}
from which the Liouvillian behavior emerges only for vanishing values of $\mbox{\boldmath $\nabla$}_{\xi} \cdot \mathbf{w}$ \cite{02A,EPL18}.

For the periodic (an)harmonic motions resumed by two-dimensional phase-space close orbits parameterized by the classical path, $\mathcal{C}\to V_{_{\mathcal{C}}}$, 
the integration of the time change of $\mathcal{W}$, $\partial \mathcal{W}/\partial\tau$, over the volume $V_{\mathcal{C}}$  results into
\begin{equation}
\int_{V_{_{\mathcal{C}}}}dV\, \frac{\partial \mathcal{W}}{\partial \tau} =
\int_{V_{_{\mathcal{C}}}}dV\, \left(\frac{D\mathcal{W}}{D\tau} - \mathbf{v}_{\xi(\mathcal{C})} \cdot \mbox{\boldmath $\nabla$}_{\xi} \mathcal{W}\right) =
\frac{D~}{D\tau}\int_{V_{_{\mathcal{C}}}}dV \,\mathcal{W} - \int_{V_{_{\mathcal{C}}}}dV \,\mbox{\boldmath $\nabla$}_{\xi}\cdot (\mathbf{v}_{\xi(\mathcal{C})}\mathcal{W}), 
\label{alexquaz51BB}
\end{equation}
where the integrated probability is identified by
\begin{equation}
\varsigma_{(\mathcal{C})} =\int_{V_{_{\mathcal{C}}}}dV\,\mathcal{W}.
\label{alexquaz60}
\end{equation}
By using the result from Eq.~(\ref{alexquaz51BB}) for evaluating the volume integration of Eq.~(\ref{alexquaz51}), after some math manipulations involving Eqs.~(\ref{alexquaz57B})-(\ref{alexquaz59}), one obtains \cite{EPL18,NovoPaper02}
\begin{equation}
 \frac{D~}{D\tau}\varsigma_{(\mathcal{C})} =\frac{D~}{D\tau}\int_{V_{_{\mathcal{C}}}}dV \,\mathcal{W} = \int_{V_{_{\mathcal{C}}}}dV \,\left[\mbox{\boldmath $\nabla$}_{\xi}\cdot (\mathbf{v}_{\xi(\mathcal{C})}\mathcal{W}) - \mbox{\boldmath $\nabla$}_{\xi}\cdot \mbox{\boldmath$\mathcal{J}$}\right],
\label{alexquaz51CC}
\end{equation}
from which, the quantum corrections are obtained from $\Delta \mbox{\boldmath$\mathcal{J}$} = \mbox{\boldmath$\mathcal{J}$} - \mathbf{v}_{\xi(\mathcal{C})}\mathcal{W}$, which is recast in the form of a path integral written as \cite{EPL18,NovoPaper02}
\begin{equation}
 \frac{D~}{D\tau}\varsigma_{(\mathcal{C})} = -\int_{V_{_{\mathcal{C}}}}dV\, \mbox{\boldmath $\nabla$}_{\xi}\cdot \Delta \mbox{\boldmath$\mathcal{J}$} = -\oint_{\mathcal{C}}d\ell\, \Delta\mbox{\boldmath$\mathcal{J}$}\cdot \mathbf{n}\equiv -\oint_{\mathcal{C}}d\ell\, \mbox{\boldmath$\mathcal{J}$}\cdot \mathbf{n}.
\label{alexquaz51DD}
\end{equation}
with the unitary vector, $\mathbf{n}$, computed from $\mathbf{n}= (-d{k}_{_{\mathcal{C}}}/d\tau, d{x}_{_{\mathcal{C}}}/d\tau) \vert\mathbf{v}_{\xi(\mathcal{C})}\vert^{-1}$, with $\mathbf{n}\cdot\mathbf{v}_{\xi(\mathcal{C})}= 0$.

One thus has a parametric integral given by
\begin{equation}
\frac{D~}{D\tau}\varsigma_{(\mathcal{C})}
\bigg{\vert}_{\tau = T} = -\oint_{\mathcal{C}}d\ell\, \Delta\mbox{\boldmath$\mathcal{J}$}\cdot \mathbf{n} = -
\int_{0}^{T}d\tau\, \Delta \mathcal{J}_p(x_{_{\mathcal{C}}}\bb{\tau},\,k_{_{\mathcal{C}}}\bb{\tau};\tau)\,\,\frac{d}{d\tau}{x}_{_{\mathcal{C}}}\bb{\tau},
\label{alexquaz51EE}
\end{equation}
where the line element, $\ell$, was written as $d\ell \equiv \vert\mathbf{v}_{\xi(\mathcal{C})}\vert d\tau$, and, in the final form, one has $x_{_{\mathcal{C}}}\bb{\tau}$ and $k_{_{\mathcal{C}}}\bb{\tau}$ as solutions for the classical Hamiltonian problem. 

As it has been anticipated, the above introduced formal developments are useful in describing the loss of information related to quantum decoherence.
Considering that the informational content of quantum systems are subjected to self-contained description of the phase-space fluid analogs in the Wigner framework, from the dimensionless version of Eq.~\eqref{pureza} for quantum purity \cite{EPL18},
\begin{eqnarray}
\mathcal{P} = 2\pi \int_{V}dV\,\, \mathcal{W}^2.
\label{alexquaz63}
\end{eqnarray}
after suitable manipulations \cite{EPL18,NovoPaper,NovoPaper02} involving Eq.~\eqref{alexquaz59}, one obtains
\footnote{In particular, if one notices that $\mbox{\boldmath $\nabla$}_{\xi}\cdot\mbox{\boldmath$\mathcal{J}$} = \mathcal{W}\,\mbox{\boldmath $\nabla$}_{\xi}\cdot\mathbf{w}+ \mathbf{w}\cdot \mbox{\boldmath $\nabla$}_{\xi}\mathcal{W}$.}
\begin{eqnarray}
\frac{1}{2\pi}\frac{D\mathcal{P}}{D\tau} &=& \frac{D~}{D\tau}\left(\int_{V}dV\, \mathcal{W}^2\right)\nonumber\\
&=&
\int_{V}dV\,\left[\frac{D~}{D\tau} \mathcal{W}^2 + \mathcal{W}^2 \mbox{\boldmath $\nabla$}_{\xi}\cdot \mathbf{v}_{\xi(\mathcal{C})}\right]\nonumber\\
&=& \int_{V}dV\,\left[\frac{\partial~}{\partial \tau}\mathcal{W}^2 + \mbox{\boldmath $\nabla$}_{\xi}\cdot(\mathbf{v}_{\xi(\mathcal{C})} \mathcal{W}^2)\right]\nonumber\\
&=& -\int_{V}dV\,\left[\mathcal{W}^2 \,\mbox{\boldmath $\nabla$}_{\xi}\cdot\mathbf{w} +
\mbox{\boldmath $\nabla$}_{\xi}\cdot(\mbox{\boldmath$\mathcal{J}$} \mathcal{W} - \mathbf{v}_{\xi(\mathcal{C})} \mathcal{W}^2)\right]\nonumber\\
&=& -\left[\int_{V}dV\,\mathcal{W}^2 \,\mbox{\boldmath $\nabla$}_{\xi}\cdot\mathbf{w}
+ \oint_{\mathcal{\,\,}}d\ell\, \mathcal{W} \left(\mbox{\boldmath$\Delta\mathcal{J}$}\cdot \mathbf{n}\right)\right],
\label{alexquaz64}
\end{eqnarray}
which, in the limit of ${V\to\infty}$, is reduced to the averaged value given by $\langle \mathcal{W} \mbox{\boldmath $\nabla$}_{\xi}\cdot\mathbf{w}\rangle$ \cite{NovoPaper} where $\langle \dots \rangle \equiv Tr_{\{x,k\}}\left[\hat{\rho}(\dots)\right]$, as reported by Refs. \cite{EPL18,NovoPaper,NovoPaper02}\footnote{It has been used that $\partial \mathcal{W}/\partial \tau = - \mbox{\boldmath $\nabla$}_{\xi}\cdot\mbox{\boldmath$\mathcal{J}$} = - \mbox{\boldmath $\nabla$}_{\xi}\cdot (\mathbf{w}\,\mathcal{W}$)}.

In particular, for periodic (an)harmonic motions driven by even parity potentials, $\mathcal{U}(x) =\mathcal{U} (-x)$, the averaged term vanishes. Thus, turning back to the expression written in terms of finite values of $V$ (cf. the last line fo Eq.~\eqref{alexquaz64}), for the classical surface ${\mathcal{C}}$ enclosing the volume $V_{_{\mathcal{C}}}$, the surface term is restored and the continuity equation is recast in the form of \cite{EPL18,NovoPaper}
\begin{eqnarray}
\frac{D~}{D\tau}\mathcal{P}_{(\mathcal{C})}
\bigg{\vert}_{\tau = T} &=& -\oint_{\mathcal{C}}d\ell\, \mathcal{W}\,\Delta\mbox{\boldmath$\mathcal{J}$}\cdot \mathbf{n} \nonumber\\&=& -
\int_{0}^{T}d\tau\, \mathcal{W}(x_{_{\mathcal{C}}}\bb{\tau},\,k_{_{\mathcal{C}}}\bb{\tau};\tau) \,\Delta \mathcal{J}_p(x_{_{\mathcal{C}}}\bb{\tau},\,k_{_{\mathcal{C}}}\bb{\tau};\tau)\,\,\frac{d~}{d\tau}{x}_{_{\mathcal{C}}}\bb{\tau},
\label{alexquaz64DD}
\end{eqnarray}
where $T=2\pi$ is the period of motion, the line element, $\ell$, has been parameterized according to $d\ell \equiv \vert\mathbf{v}_{\xi(\mathcal{C})}\vert d\tau$, for $x_{_{\mathcal{C}}}\bb{\tau}$ and $k_{_{\mathcal{C}}}\bb{\tau}$ described as solutions of the classical Hamiltonian problem.

As it has been reported \cite{JCAP18,EPL18,NovoPaper,Eplus19}, the above obtained quantum purity continuity equation works as an optimized quantifier for non-classicality as well as for the flux of information encoded by the quantum system.

For this reason, in the context of a preliminarily approach for the understanding of phase-space classical and quantum properties of quantum ensembles, the quantum purity is relevant in probing the role of finite temperature and interaction parameters in determining the level of decoherence in quantum systems.

\section{Wigner function and quantum purity for the singular oscillator}

\hspace{1em}
To quantitatively interpret the distortion induced by the contribution of an inverse square potential onto the harmonic quantum background, one considers the dimensionless anharmonic Weyl transformed Hamiltonian fully described by
\begin{equation}\label{qua14}
\mathcal{H}(k,\,x) = \frac{1}{2}\left\{k^2+ x^{2} + \frac{4 \alpha^2 -1}{4 x^{2}}-2\alpha \right\},
\end{equation}
where the $\alpha$ parameter drives the anharmonic behavior, which is reduced to the harmonic oscillator dynamics when $\alpha = -1/2$. In fact, a simplifying analysis of the problem can be achieved through the Schr\"odinger equation for the quadratic plus inverse square potential, written as 
\begin{equation}\label{qua16}
\mathcal{H} \phi^{\alpha}_n(x) = \frac{1}{2}\left\{-\frac{d^{2}}{dx^{2}}+ x^{2} + \frac{4 \alpha^2 -1}{4 x^{2}}-2\alpha \right\} \phi^{\alpha}_n(x) = \varepsilon_n\,\phi^{\alpha}_n({x}),
\end{equation}
where one has identified $k\equiv -i\,(d/dx)$, $\alpha$ is a continuous value parameter, and the self-energy is given by $E_n =\hbar \omega\, \varepsilon_n$, with $\varepsilon_n = 2n + 1$ and $n$ assuming only integer values.
The corresponding wave function, solution of the Schr\"odinger Eq.~\eqref{qua16}, is written as
\begin{equation}
\phi^{\alpha}_n(x) = 2^{{1}/{2}}\,\Theta(x) \, N_n^{(\alpha)}\, x^{\alpha + \frac{1}{2}}\,\exp(-x^2/2)\,L^{\alpha}_n(x^2),
\label{qua19}
\end{equation}
where $L^{\alpha}_n$ are the {\em associated Laguerre polynomials}, and $N_n^{(\alpha)}$ is the normalization constant given by
\begin{equation}
N_n^{(\alpha)} = \sqrt{\frac{n!}{\Gamma(n+\alpha+1)}}, 
\label{qua20}
\end{equation}
where $\Gamma(n) = (n-1)!$ is the {\em gamma function} and, finally, $\Theta(x)$ is the {\em heavyside function} introduced to delimit the solution to the interval of $0 < x < \infty$.

From Eq.~\eqref{qua19}, the dimensionless form of the Wigner function (cf. Eq.~\eqref{alexDimW}) related to $\phi^{\alpha}_n(x)$ results into
\small\begin{eqnarray}\label{DimW2}
\mathcal{W}_n^{\alpha}(x, \, k) 
&=&2 (N_n^{(\alpha)})^2\, \pi^{-1} \int^{+\infty}_{-\infty} \hspace{-.15cm}dy\,\Theta(x+y)\Theta(x-y)\,(x^2-y^2)^{\frac{1}{2}+\alpha}\, \exp\left(2\,i\, k\,y\right)\\
&&\qquad\qquad\qquad\qquad\qquad\qquad\qquad\exp\left[-(x^2+y^2)\right] L_n^{\alpha}\left((x+y)^2\right)\,L_n^{\alpha}\left((x-y)^2\right)
\nonumber\\
&=& \frac{2}{\pi} \int^{+x}_{-x} \hspace{-.15cm}dy\,\exp\left(2\,i\, k\,y\right)\,\exp\left[-(x^2+y^2)\right] \,\sum_{j=0}^n \frac{L_{n-j}^{\alpha+2j}\left(2 (x^2+y^2)\right) }{\Gamma(\alpha+j+1)}
\frac{(x^2-y^2)^{\frac{1}{2}+\alpha+2j}}{j!},\nonumber
\end{eqnarray}\normalsize
where it has been noticed that \cite{NovoPaper}
\begin{equation}
L_n^{\alpha}\left(x\right)\,L_n^{\alpha}\left(y\right) = \frac{\Gamma(n+\alpha+1)}{n!}\,\sum_{j=0}^n \frac{L_{n-j}^{\alpha+2j}\left(x+y\right) }{\Gamma(\alpha+j+1)}
\frac{x^j y^j}{j!}.
\end{equation}

\subsection{Computation of purity - An example of pure state}

\hspace{1em} A time-dependent quasi-Gaussian pure state can be constructed from the quantum superposition given by
\begin{equation}
\label{supp}
\mathcal{G}_{\alpha}(x,\,\tau_{}) =\mathcal{N}_{_F} \sum_{n=0}^{\infty}c^{\alpha}_n(\tau_{})\, \phi^{\alpha}_n(x),
\end{equation}
with $c^{\alpha}_n$ identified by $u^n\,N^{-1}_n(\alpha)\,\exp(- (i/2) \tau_{})$, with $u\equiv u(\gamma,\,\tau_{}) = \exp(-\gamma + i \,\tau_{})$, where $\gamma$ is an arbitrary weight parameter which constrains the expansion coefficients to $|u| < 1$, so as to have
\begin{eqnarray}
\mathcal{G}_{\alpha}(x,\,\tau_{}) 
&=& \mathcal{N}_{_F}\,\Theta(x)\, x^{\alpha + \frac{1}{2}}\,\exp(-x^2/2)\,\sum_{n=0}^{\infty}u^n\,L^{\alpha}_n(x^2)\nonumber\\
&=& \mathcal{N}_{_F}\,\Theta(x)\, x^{\alpha + \frac{1}{2}}\,(1-u)^{-(1+\alpha)}\exp\left[-\frac{1}{2}\left(\frac{1+u}{1-u}\right)x^2\right]
\label{eqn22},
\end{eqnarray}
with 
\begin{eqnarray}
\mathcal{N}_{_F} &=&\left[\frac{(1- e^{-2\gamma})^{1+\alpha}}{2\Gamma(1+\alpha)}\right]^{{1}/{2}},
\end{eqnarray}
and from which the quasi-Gaussian profile is identified by
\begin{eqnarray}
\vert\mathcal{G}_{\alpha}(x,\,\tau_{})\vert^2 
&=& \frac{2\,\chi _{(\gamma,\tau_{})}^{1+\alpha}}{\Gamma(1+\alpha)}\,\Theta(x)\,x^{1+2\alpha}\, \exp(-\chi _{(\gamma,\tau_{})} x^2),
\label{eqnstart}\end{eqnarray}
with
\begin{equation}
\chi _{(\gamma,\tau_{})}= \frac{\sinh(\gamma)}{\cosh(\gamma)-\cos(\tau_{})}.
\label{eqn23}
\end{equation}
The above result has been discussed, for instance, in phase-space quantum cosmological scenarios in order to reproduce the classical cosmological background from a quantum approach \cite{JCAP18} and can be relevant in more general contexts involving the discussion of the informational content of cosmological scenarios \cite{AEBRDR}.
By substituting the expression from Eq.~(\ref{eqn22}) into Eq.~(\ref{alexDimW}) one obtains \cite{JCAP18}\footnote{For semi-integer values of $\alpha$, one has the finite series
\begin{equation}
(1-s^2)^{\frac{1}{2}+\alpha} = (1-s^2)^{1 + \nu} = \sum_{j=0}^{1 + \nu}
(-1)^j\,\frac{s^{2j}(1+\nu)!}{j!(1+\nu-j)!} = \sum_{j=0}^{1 + \nu}
\frac{\Gamma(3/2+\alpha)}{\Gamma(3/2+\alpha-j)\Gamma(1+j)},
\end{equation}
where $\alpha$ has been rewritten as $\alpha = 1/2 + \nu$, with $\nu = 0,\,1,\,2,\, \dots$.
By substituting the above expression into Eq.~(\ref{eqn27}), one obtains
\begin{eqnarray}
\label{finalform2}
\mathcal{W}^{\alpha}(x,\,k;\,\tau_{})
&=& \frac{4}{\pi}\frac{\Gamma(3/2+\alpha)}{\Gamma(1+\alpha)}
(\chi \,x^2)^{1+\alpha}
\exp\left(-\chi \,x^2\right)\sum_{j=0}^{\frac{1}{2}+\alpha}\frac{(-1)^j}{\Gamma(1+j)\,\Gamma(3/2+j+\alpha)}\,
\,\nonumber\\
&&\qquad \int_{0}^{+1}ds\, s^{2j}\,\exp\left(-\chi \,x^2\,s^2\right)\,\cos\left(2\,x\,s (k + \tilde{\chi }\, x)\right)\nonumber\\
&=& \frac{1}{\sqrt{\pi}}\frac{\Gamma(3/2+\alpha)}{\Gamma(1+\alpha)}
(\chi \,x^2)^{1+\alpha}
\exp\left(-\chi \,x^2\right)
\,\sum_{j=0}^{\frac{1}{2}+\alpha}\frac{x^{-(1+2j)}}{\Gamma(1+j)\,\Gamma(3/2+j+\alpha)}\nonumber\\
&&\qquad \,
\frac{d^j}{d\chi^j}\left[\chi^{-{1}/{2}} \exp\left[-\frac{(k+\tilde{\chi }x)^{2}}{\chi }\right]
\left(\mbox{Erf}[\zeta(\chi ,\tilde{\chi })] + h.c.\right)
\right],
\end{eqnarray}
with $\zeta(\chi ,\tilde{\chi }) = \chi^{{1}/{2}}(x + i\,\chi^{-1} (k+\tilde{\chi }x))$, and where the subindex $_{(\gamma,\tau_{})}$ has been suppressed.}
\begin{eqnarray}
\label{eqn27}
\mathcal{W}^{\alpha}(x,\,k;\,\tau_{}) &=& \frac{2\,\chi _{(\gamma,\tau_{})}^{1+\alpha}}{\pi\,\Gamma(1+\alpha)}\,\int_{-\infty}^{+\infty}dy\,\Theta(x+y)\Theta(x-y)\,(x^2-y^2)^{\frac{1}{2}+\alpha}\nonumber\\
&&\qquad\qquad\qquad \exp\left(-\chi _{(\gamma,\tau_{})} (x^2+y^2)\right)\,\exp\left(2\,i\,y(k + \tilde{\chi }_{(\gamma,\tau_{})}\, x)\right)\\
&=& \frac{2\,\chi _{(\gamma,\tau_{})}^{1+\alpha}}{\pi\,\Gamma(1+\alpha)}\,x^{2(1+\alpha)}
\exp\left(-\chi _{(\gamma,\tau_{})} x^2\right)\,\nonumber\\
&&\qquad \int_{-1}^{+1}ds\, (1-s^2)^{\frac{1}{2}+\alpha}\, \exp\left(-\chi _{(\gamma,\tau_{})}\,x^2\,s^2\right)\,\exp\left(2\,i\,x\,s (k + \tilde{\chi }_{(\gamma,\tau)}\, x)\right),\nonumber
\end{eqnarray}
where $y$ has been parameterized by $ y = x\,s$ and, besides $\chi _{(\gamma,\tau_{})}$ from Eq.~\eqref{eqn23}, one has
\begin{equation}
\tilde{\chi }_{(\gamma,\tau_{})} = -\frac{\sin(\tau_{})}{\cosh(\gamma)-\cos(\tau_{})}.
\label{tilde}
\end{equation}
By identifying the integrand even and odd parity components before introducing the power series expansion,
\begin{equation}
\cos(2r) = \sum_{j=0}^{\infty}(-1)^j\,\frac{2^{2j}}{(2j)!}r^{2k}, 
\end{equation}
for the even parity contribution from the exponential, one can evaluate the following integration contributions,
\begin{eqnarray}
\lefteqn{2\int_{0}^{+1}ds\, (1-s^2)^{\frac{1}{2}+\alpha}\, s^{2j}\,\exp\left(-\chi _{(\gamma,\tau_{})}\,x^2\,s^2\right) =}\nonumber\\
&&\qquad\qquad
\Gamma(3/2+\alpha)\Gamma(1/2+j)\,\, _1\mathcal{F}_1(1/2+j,2+\alpha+j,-\chi _{(\gamma,\tau_{})}\,x^2\,s^2),
\end{eqnarray}
where $_1\mathcal{F}_1$ is the (first kind) confluent hypergeometric function. 

The reduced form of the Wigner function is thus given by \begin{eqnarray}
\label{finalform}
\mathcal{W}^{\alpha}(x,\,k;\,\tau_{}) &=& \frac{2}{\sqrt{\pi}}\frac{\Gamma(3/2+\alpha)}{\Gamma(1+\alpha)}
(\chi _{(\gamma,\tau_{})}x^2)^{1+\alpha}
\exp\left(-\chi _{(\gamma,\tau_{})} \,x^2\right)\\
&&\qquad
\sum_{j=0}^{\infty}\frac{(-1)^j\,(k\,x + \tilde{\chi }_{(\gamma,\tau)}\, x^{2})^{2j}}{\Gamma(1+j)\,\Gamma(2+j+\alpha)}\,\,_1\mathcal{F}_1(1/2+j,2+\alpha+j,-\chi _{(\gamma,\tau_{})}\,x^2\,s^2).\nonumber
\end{eqnarray}

Before verifying the purity of the {\em quasi}-Gaussian quantum state from Eq.~(\ref{eqnstart}), its normalization condition is recovered from observing that, from Eq.~(\ref{eqn27}), one has
\begin{eqnarray}
\int_{0}^{\infty}\hspace{-.3 cm}dx\int_{-\infty}^{+\infty}\hspace{-.3 cm}dp \,\mathcal{W}^{\alpha}(x,\,k;\,\tau_{}) &=& \frac{2\,\chi^{1+\alpha}}{\pi\,\Gamma(1+\alpha)}\,\int_{0}^{\infty}\hspace{-.3 cm}dx \,x^{2(1+\alpha)}\,\exp\left(-\chi \, x^2\right)\,\int_{-\infty}^{+\infty}\hspace{-.3 cm}dk
\exp\left(2\,i\,x\,k\,s \right)\times\nonumber\\
&& \qquad\quad\int_{-1}^{+1} ds\,(1-s^2)^{\frac{1}{2}+\alpha}\, \exp\left(-\chi \, x^2\,s^2\right) \exp\left(2i\,\tilde{\chi }\, x^2\,s\right).\quad
\label{eqnA01}
\end{eqnarray}
Given that
\begin{eqnarray}
\int_{-\infty}^{+\infty}\hspace{-.3 cm}dk
\,\exp\left(2\,i\,x\,k\,s \right) &=& 2\pi\,\delta(2\,x\,s) = \frac{\pi}{\vert x\vert}\delta(s)
\label{eqnA02}
\end{eqnarray}
can be replaced into Eq.~(\ref{eqnA01}), a subsequent integration over $s$ yields
\begin{equation}
\frac{\pi}{\vert x\vert}\int_{-1}^{+1} ds\,\delta(s) (1-s^2)^{\frac{1}{2}+\alpha}\, \exp\left(-\chi \, x^2\,s^2\right) \exp\left(2i\,\tilde{\chi }\, x^2\,s\right) = \frac{\pi}{\vert x\vert}
\label{eqnA01B},
\end{equation}
as to give
\begin{eqnarray}
\int_{0}^{\infty}\hspace{-.3 cm}dx\int_{-\infty}^{+\infty}\hspace{-.3 cm}dk \,\mathcal{W}^{\alpha}(x,\,k;\,\tau_{}) &=& \frac{2\,\chi^{1+\alpha}}{\Gamma(1+\alpha)}\,\int_{0}^{\infty}\hspace{-.3 cm}dx \,x^{(1+2\alpha)}\,\exp\left(-\chi \, x^2\right) = 1\label{eqnA03}.
\end{eqnarray}

For the quantum purity (cf. Eq.~(\ref{pureza})),
\begin{equation}
\label{purepure}\mathcal{P} = 2\pi\int_{0}^{\infty}\hspace{-.3 cm}dx\int_{-\infty}^{+\infty}\hspace{-.3 cm}dk \,\left(\mathcal{W}^{\alpha}(x,\,k;\,\tau_{})\right)^{2},
\end{equation}
a similar strategy would lead to \cite{JCAP18}
\begin{eqnarray}
\lefteqn{\int_{0}^{\infty}\hspace{-.3 cm}dx\int_{-\infty}^{+\infty}\hspace{-.3 cm}dk \,\left(\mathcal{W}^{\alpha}(x,\,k;\,\tau_{})\right)^{2} =\frac{4\,\chi^{2(1+\alpha)}}{\pi^2\,\Gamma^2(1+\alpha)}\times}\nonumber\\
&& \,\int_{0}^{\infty}\hspace{-.3 cm}dx \,x^{4(1+\alpha)}\,\exp\left(-2\,\chi \, x^2\right)
\,\int_{-\infty}^{+\infty}\hspace{-.3 cm}dk
\,\exp\left(2\,i\,x\,k \,(s+r) \right)\times
\nonumber\\
&& \qquad\int_{-1}^{+1} ds\int_{-1}^{+1} dr\,[(1-r^2)(1-s^2)]^{\frac{1}{2}+\alpha}\, \exp\left(-\chi \, x^2\,(r^2+s^2)\right) \exp\left(2i\,\tilde{\chi }\, x^2\,(s+r)\right). \quad\,
\label{eqnA04}
\end{eqnarray}
Again, from Eq.~\eqref{eqnA02} -- with $s$ replaced by $s + r$ --
substituted into Eq.~(\ref{eqnA04}), followed by an integration over the variable $r$, one obtains
\begin{eqnarray}
\lefteqn{\int_{0}^{\infty}\hspace{-.3 cm}dx\int_{-\infty}^{+\infty}\hspace{-.3 cm}dk \,\left(\mathcal{W}^{\alpha}(x,\,k;\,\tau_{})\right)^{2} = \frac{4\,\chi^{2(1+\alpha)}}{\pi^2\,\Gamma^2(1+\alpha)}\times}\nonumber\\ &&\qquad\qquad\int_{0}^{\infty}\hspace{-.3 cm}dx \,x^{(3+4\alpha)}\,\exp\left(-2\,\chi \, x^2\right)\,\int_{-1}^{+1} ds\,(1-s^2)^{1+2\alpha}\, \exp\left(-2\,\chi \, x^2\,s^2\right).
\label{eqnA05}
\end{eqnarray}
Through the evaluation of the integral over $x$, one finally obtains
\begin{eqnarray}
\int_{0}^{\infty}\hspace{-.3 cm}dx \,x^{(3+4\alpha)}\,\exp\left(-2\,\chi \, x^2\,(1+s^2)\right) = \frac{1}{2^{3+2\alpha}\chi^{2(1+\alpha)}}
\frac{\Gamma(2(1+\alpha))}{(1+s^2)^{2(1+\alpha)}},
\label{eqnA06}
\end{eqnarray}
which, once substituted into Eq~(\ref{eqnA05}), returns
\begin{eqnarray}
\int_{0}^{\infty}\hspace{-.3 cm}dx\int_{-\infty}^{+\infty}\hspace{-.3 cm}dp \,\left(\mathcal{W}^{\alpha}(x,\,k;\,\tau_{})\right)^{2}
 &=&
\frac{1}{2^{1+2\alpha}\pi}
\frac{\Gamma(2(1+\alpha))}{\Gamma^2(1+\alpha)} \int_{-1}^{+1} ds\,\frac{(1-s^2)^{1+2\alpha}}{(1+s^2)^{2(1+\alpha)}}\nonumber\\
 &=&
\frac{1}{2^{2+2\alpha}\pi}
\frac{\Gamma(2(1+\alpha))}{\Gamma^2(1+\alpha)} \frac{\sqrt{\pi} \Gamma(1+\alpha)}{2\Gamma(3/2+\alpha)}
\nonumber\\
 &=&\frac{1}{2\pi},
\label{eqnA07}
\end{eqnarray}
which shows that the quantum state defined by Eq.~(\ref{eqn27}) is a pure one.

\section{Wigner quantum ensemble}

\hspace{1em} One now turns the attention to the construction of the canonical ensemble from the Weyl-Wigner framework.
For a time independent Hamiltonian, $\mathcal{H}$, the quantum propagator (Green's function) can be written as
\begin{equation}
\Delta(x,\tau;\,x^{\prime},0) = \langle x \vert \exp(-i\,\tau\, \hat{\mathcal{H}}) \vert x^{\prime} \rangle,
\end{equation}
where, by replacing $\tau$ by $- i \, \beta \hbar\omega$, where the inverse of $\beta$ is the dimensionless product of the temperature, $\mathcal{T}$, by the Boltzmann constant, $k_{B}$, i.e. $\beta^{-1} = k_{B} \mathcal{T}$, one has
\begin{equation}
\Delta^{\alpha}(x,- i \, \beta;\,x^{\prime},0) = \sum_n \exp(-\beta\hbar\omega\, \varepsilon_n)\,\phi^{*\alpha}_n({x})\phi^{\alpha}_n({x^{\prime}})
= \rho^{\alpha}(x,\,x^{\prime}),
\end{equation}
where the additional index, $\alpha$, has been introduced in order to set the correspondence with Eq.~\eqref{qua16}, i.e. $\mathcal{H}\,\phi^{\alpha}_n({x}) = \varepsilon_n\,\phi^{\alpha}_n({x})$.

The functional $\rho^{\alpha}(x,\,x^{\prime};\,\beta)$ is identified as the thermal density matrix in the coordinate representation of the state operator for the canonical ensemble of a system in equilibrium with a heat reservoir at the temperature $\beta$.
Once that $\rho^{\alpha}(x,\,x^{\prime};\,\beta)$ is identified with a density matrix operator $\hat{\rho}$, their delocalization aspects can be parameterized by the displacement relations, $x \to x+y$ and $x^{\prime} \to x - y$, so as to identify the Fourier transform of the thermal density matrix with the thermalized phase-space probability distribution,
\begin{eqnarray}
\label{alexDimW222}
\Omega^{\alpha}(x, \, k;\,\beta) &=& \pi^{-1} \int^{+\infty}_{-\infty} \hspace{-.15cm}dy\,\exp{\left(2\, i \, k \,y\right)}\,\rho^{\alpha}(x+y,\, x-y;\,\beta).
\end{eqnarray}
According to the matrix operator representation, one has a correspondence given by $\rho^{\alpha}(x+y,\, x-y;\,\beta) \to \mathcal{Z}^{-1}\exp(-\beta\hbar\omega\, \hat{\mathcal{H}}) \equiv \mathcal{Z}^{-1}\hat{\Omega}^{\alpha}$, where the unnormalized density matrix, $\hat{\Omega}^{\alpha}$, satisfies the Bloch equation, 
\begin{equation}
\frac{\partial \hat{\Omega}^{\alpha}}{\partial \beta} = - \hat{\mathcal{H}}\hat{\Omega}^{\alpha} = -\hat{\Omega}^{\alpha}\hat{\mathcal{H}},
\end{equation}
with $\hat{\Omega}^{\alpha}(\beta=0) \propto \mathbb{I}$, and from which the starting point for more systematic calculations in quantum statistical mechanics follows from the definition of the partition function as a trace related functional, $\mathcal{Z}\equiv\mathcal{Z}(\beta) = Tr[\exp(-\beta\hbar\omega\, \hat{\mathcal{H}})]$.
In fact, from Eq.~\eqref{alexDimW222}, the corresponding (normalized) Wigner function can be written as
\begin{eqnarray}
\label{alexDimW222B}
\mathcal{W}_\Omega^{\alpha}(x, \, k;\,\beta) &=& (\mathcal{Z}\,\pi)^{-1} \int^{+\infty}_{-\infty} \hspace{-.15cm}dy\,\exp{\left(2\, i \, k \,y\right)}\,\rho(x+y,\, x-y;\,\beta),
\end{eqnarray}
with
\begin{equation}
\mathcal{Z}(\beta) = Tr\left[\exp(-\beta\hbar\omega\, \hat{\mathcal{H}})\right] = \int^{+\infty}_{-\infty} \hspace{-.15cm}dx\,\int^{+\infty}_{-\infty} \hspace{-.15cm}dk\,\Omega^{\alpha}(x, \, k;\,\beta).
\end{equation}

Finally, for the {\em singular oscillator} driven by the Hamiltonian from Eq.~\eqref{qua16}, the canonical ensemble of quantum states is described in terms of the Wigner distribution written as
\begin{eqnarray}
\label{alexDimW222B}
\mathcal{W}_\Omega^{\alpha}(x, \, k;\,\beta) &=&\frac{\exp(\beta\hbar\omega)}{\mathcal{Z}(\beta)}\sum_{n=0}^{\infty} \mathcal{W}_n^{\alpha}(x, \, k)\, \exp(-2n\,\beta\hbar\omega),\end{eqnarray}
with $\mathcal{W}_n^{\alpha}$ from Eq.~\eqref{DimW2}, which leads to
\small\begin{eqnarray}
\mathcal{W}_\Omega^{\alpha}(x, \, k;\,\beta) &=& \frac{2 \exp(-{\beta\hbar\omega})}{\mathcal{Z}(\beta)\,\pi}
\int^{+x}_{-x}
\hspace{-.15cm}dy\,\exp\left(2\,i\, k\,y\right)\,\exp\left[-(x^2+y^2)\right]\,(x^2-y^2)^{\frac{1}{2}+\alpha}\times\\
&&\qquad\qquad\qquad
\sum_{n=0}^{\infty}\bigg{\{}\exp(-2n\,\beta\hbar\omega)\frac{n!}{\Gamma(\alpha+n+1)}L_n^{\alpha}\left((x+y)^2\right)\,L_n^{\alpha}\left((x-y)^2\right)\bigg{\}}.\quad\quad\quad
\nonumber
\end{eqnarray}\normalsize
From the properties of the {\em associated Laguerre polynomials} \cite{Gradshteyn}, one has
\begin{eqnarray}
\label{alexDimW222C}
\lefteqn{\sum_{n=0}^{\infty}\bigg{\{}\exp(-2n\,\beta\hbar\omega)\frac{n!}{\Gamma(\alpha+n+1)}L_n^{\alpha}\left((x+y)^2\right)\,L_n^{\alpha}\left((x-y)^2\right)\bigg{\}} =} \\&&\quad\quad\quad\quad\quad\quad\quad\quad\quad\quad\quad\quad\frac{(x^2-y^2)^{-\alpha}}{(1-\lambda)\lambda^{\frac{\alpha}{2}}}
\exp\left[-\frac{2\lambda}{1-\lambda}(x^2+y^2)\right] \mathcal{I}_{\alpha} \left(\frac{2\lambda^{\frac{1}{2}}}{1-\lambda}(x^2-y^2)\right),\quad\quad\quad\nonumber
\nonumber
\end{eqnarray}\normalsize
with $\lambda = \exp(-{2\beta\hbar\omega})$, where $\mathcal{I}_{\alpha}$ is the {\em modified Bessel function of the first kind}, and which can be substituted into Eq.~\eqref{alexDimW222B} as to give
\small\begin{eqnarray}\label{finalwigner}
\mathcal{W}_\Omega^{\alpha}(x, \, k;\,\beta) &=& \frac{\exp({\alpha\beta\hbar\omega})}{\sinh(\beta\hbar\omega)\mathcal{Z}(\beta)\,\pi}\times\\&&\qquad
\int^{+x}_{-x}
\hspace{-.15cm}dy\,\exp\left(2\,i\, k\,y\right)\,(x^2-y^2)^{\frac{1}{2}}
\exp\left[-\coth(\beta\hbar\omega)(x^2+y^2)\right]\,\mathcal{I}_{\alpha} \left(\frac{x^2-y^2}{\sinh(\beta\hbar\omega)}\right).\nonumber
\end{eqnarray}\normalsize

Of course, from the normalization of each contribution, $\mathcal{W}_n^{\alpha}$, for the series expansion (\ref{alexDimW222B}), it is possible to verify that 
\begin{eqnarray}
\mathcal{Z}(\beta) =\exp(-\,\beta\hbar\omega)\sum_{n=0}^{\infty}\exp(-2n\,\beta\hbar\omega) = \frac{\exp(-\,\beta\hbar\omega)}{1- \exp(-2\,\beta\hbar\omega)} = \frac{1}{2\sinh(\beta\hbar\omega)},
\end{eqnarray}
which does not depend on the anharmonic distortion driven by $\alpha$, and describes an equivalent thermodynamics to that one of a $1$-dim harmonic oscillator with characteristic frequency given by $2 \omega$, as it has been pointed out in the begging of our analysis. The above result can also be consistently reproduced from a direct phase-space volume integration of the complete expression for $\mathcal{W}_\Omega^{\alpha}(x, \, k;\,\beta) $ from Eq.~(\ref{finalwigner}).

Thus, the question to be posed here is if the elements for distinguishing such quantum systems are provided from the definition of quantum purity (cf. Eq.~\eqref{purepure}). 
The Eq.~\eqref{finalwigner} thus yields
\small\begin{eqnarray}\label{finalwigner22}
\mathcal{P}^{\alpha}(\beta) &=& \frac{8\exp({2\alpha\beta\hbar\omega})}{\pi}
\int_{0}^{\infty}\hspace{-.3 cm}dx\,
\int^{+x}_{-x}\hspace{-.15cm} dz\,
\int^{+x}_{-x}\hspace{-.15cm}dy\,
\int_{-\infty}^{+\infty}\hspace{-.3 cm}dk \exp\left(2\,i\, k\,(y+z)\right)\times\\
&&\qquad\qquad
\left[(x^2-y^2)(x^2-z^2)\right]^{\frac{1}{2}}
\exp\left[-\coth(\beta\hbar\omega)(2x^2+y^2+z^2)\right]\times\nonumber\\
&&\qquad\qquad\qquad\qquad\qquad\qquad\,\mathcal{I}_{\alpha} \left(\frac{x^2-y^2}{\sinh(\beta\hbar\omega)}\right)
\,\mathcal{I}_{\alpha} \left(\frac{x^2-z^2}{\sinh(\beta\hbar\omega)}\right)\nonumber\\
&=& 8\exp({2\alpha\beta\hbar\omega})
\int_{0}^{\infty}\hspace{-.3 cm}dx\,
\int^{+x}_{-x}\hspace{-.15cm} dz\,(x^2-z^2)\exp\left[-2\coth(\beta\hbar\omega)(x^2+z^2)\right]
\mathcal{I}^{2}_{\alpha} \left(\frac{x^2-z^2}{\sinh(\beta\hbar\omega)}\right)\nonumber\\
&=& 8\exp({2\alpha\beta\hbar\omega})
\int^{+1}_{-1}\hspace{-.15cm} ds\,(1-s^2)
\int_{0}^{\infty}\hspace{-.3 cm}dx\,x^3
\exp\left[-2x^2\coth(\beta\hbar\omega)(1+s^2)\right]
\mathcal{I}^{2}_{\alpha} \left(x^2\frac{1-s^2}{\sinh(\beta\hbar\omega)}\right)\quad\nonumber\\
&=& 4\exp({2\alpha\beta\hbar\omega})
\int^{+1}_{-1}\hspace{-.15cm} ds\,(1-s^2)
\int_{0}^{\infty}\hspace{-.3 cm}d\mathcal{X}\,\mathcal{X}
\exp\left[-2\mathcal{X}\coth(\beta\hbar\omega)(1+s^2)\right]
\mathcal{I}^{2}_{\alpha} \left(\mathcal{X}\frac{1-s^2}{\sinh(\beta\hbar\omega)}\right),\quad\nonumber
\end{eqnarray}\normalsize
which, for semi-integer values of $\alpha$, results into
\small\begin{eqnarray}\label{finalwigner2}
\mathcal{P}^{\alpha}(\beta) &=&\frac{1}{2^{2\alpha-1}\sqrt{\pi}}\frac{\Gamma(\alpha+3/2)}{\Gamma(\alpha+1)}
\exp({2\alpha\beta\hbar\omega})\,\tanh^{2}(\beta\hbar\omega)\,\mbox{sech}^{2\alpha}(\beta\hbar\omega)
\times\\
&&\qquad\qquad
\int^{+1}_{-1}\hspace{-.15cm} ds\,\frac{(1 - s^2)^{2\alpha+1}}{(1 + s^2)^{2\alpha+2}}\,
_2\mathcal{F}_1\left[\alpha+1/2,\,\alpha+3/2,\,2\alpha+1,\,\left(\frac{1 - s^2}{1 + s^2}\mbox{sech}(\beta\hbar\omega)\right)^2
\right]
\end{eqnarray}\normalsize
where $_2\mathcal{F}_1$ is the {\em ordinary hypergeometric function}, and which, after extenuating manipulations (or even numerically, from Eq.~\eqref{finalwigner22}), results into $\mathcal{P}^{\alpha}(\beta) = \tanh(\beta\hbar\omega)$, which has an obvious and consistent interpretation.
For the limit of higher temperatures ($\beta \to 0$), one has maximal statistical mixtures, with $\mathcal{P}^{\alpha}(0) =0$. For the limit of lower temperatures ($\beta \to \infty$), 
from Eq.~\eqref{finalwigner2}, the integration for $\mathcal{P}^{\alpha}(\beta)$ can be evaluated analytically, 
 \begin{eqnarray}
 \label{finalwigner3}
\lim_{\beta\to\infty}\mathcal{P}^{\alpha}(\beta) &=& 
\frac{4}{\sqrt{\pi}}
\frac{\Gamma(\alpha+3/2)}{\Gamma(\alpha+1)}
\int^{+1}_{-1}\hspace{-.15cm} ds\,
\frac{(1 - s^2)^{2\alpha+1}}{(1 + s^2)^{2\alpha+2}} = 1,
\end{eqnarray}
and one recovers the pure state (zero temperature) configuration, which corresponds to the ground state Wigner function.
The final results do not depend on $\alpha$.
As it can be depicted from Eq.~\eqref{finalwigner}, despite the obvious dependence of the Wigner functions on the interaction parameter, $\alpha$, it does not affect the decoherence pattern for the quantum ensemble.

\section{Discussion and Conclusions}

\hspace{1em}
In this article, the quantum purity related to the thermalized Wigner function of an anharmonic quantum system has been obtained and discussed in the context of obtaining quantum information quantifiers for measuring decoherence effects of quantum ensembles.
Stable quantum configurations, according to finite temperature and interaction parameter values have been identified as to quantify the level of mixing and decoherence of quantum ensembles.
The evinced non-linear deviation from the harmonic oscillator profile, and its corresponding phase-space profile supported by Wigner functions, for a pure state and a statistical mixture (quantum ensemble) have been discussed in an overall analytical context.

The {\em singular oscillator} quantum Hamiltonian discussed here exhibits an already known systematic analytical appeal -- similar to those ones of the $1$-dim reduction of the $3$-dim Schr\"odinger equation \cite{NovoPaper} -- also identified in the context of hyperbolic quantum wells \cite{EPL18} and in the implementation of typical scenarios of quantum cosmology \cite{JCAP18}.

In the context of a preliminarily approach for the understanding of phase-space classical and quantum correspondence, the quantum purity was obtained. The most relevant result of our investigation is concerned with the role of the quantum purity in depicting equivalent measurable information from inequivalent encoding mechanisms (driven by the $\alpha$-term interactions) as described by an equivalent $1$-dim harmonic oscillator, on one hand, and by the {\em singular oscillator}, on the other hand.
Essentially, it is related to the fact that both quantum systems have the thermodynamic information content described by equivalent partition functions which, in case of singular oscillators, also are not affected by the $\alpha$-term interactions.

Concerning some further investigations, the fluid analog of the phase-space information flux discussed here can indeed be extended to the description of von Neumann and R\'enyi entropies in the context of statistical ensembles. The system here considered supports periodic motions which delineate an enclosing phase-space classical path from which quantum distortions can be identified \cite{EPL18,NovoPaper02}.
Therefore, it comprises the initial setup for employing the Wigner flow framework to discuss how quantum regimes are far from classical ones in the scenario of statistical mixtures.

To conclude, it is relevant to emphasize that the {\em singular oscillator} quantum Hamiltonian also appears in the context of quantum to classical transition in the Ho\v{r}ava-Lifshitz quantum cosmology \cite{JCAP18} through the Wheeler-DeWitt (WdW) equation for a wave function of the Universe in a parametric form given by
\begin{equation}\label{eqn16}
\left\{\frac{d^{2}}{d x^{2}} - x^{2} - \frac{4 \alpha^2 -1}{4 x^{2}} + 2 (\alpha+ 2n + 1) \right\} \psi^{\alpha}_n(a\bb{x})=0,
\end{equation}
with the Friedmann scale parameter identified by $a = g_{_{C}}^{-{1}/{4}} x$, and where one identifies the associated ``energy'' eigenvalues with $E_n =g_{_{C}}^{{1}/{2}}\left(2n + 1\right)$, with the curvature parameter, $g_{_{C}}$, in energy unities expressed by $g_{_{C}} \equiv  \ell^2\,c^4/G^2 \equiv \mathcal{M}^2\,c^4 \equiv \hbar^2\omega^2$, where $\ell$ and $\mathcal{M}$ are, respectively, length and mass scales associated to the net gravitational effect which eventually produces the (residual or even null) curvature of the Universe\footnote{For instance, for $\ell$ identified with the Planck length, $\ell_{_{Pl}}$, one has $\mathcal{M}$ identified with the Planck mass, $\mathcal{M}_{_{Pl}}$.}. It results into a cosmological ensemble purity given by
$\mathcal{P}^{\alpha}(\mathcal{T}) = \tanh(\mathcal{M}\,c^2/k_B\mathcal{T})$ such that $\mathcal{M}\,c^2 \ll k_B\mathcal{T} \sim 0.235\, meV$ in order to keep the Universe's decoherence profile consistent with its flatness, an issue which surely deserves further investigation.

{\em Acknowledgments} -- This work was supported by the Brazilian agencies FAPESP (grant 2018/03960-9) and CNPq (grant 300831/2016-1).\\

\end{document}